\documentclass[reprint,amsmath,amssymb,aps,prl]{revtex4-1}
\usepackage[colorlinks=true,linkcolor=blue]{hyperref}
\usepackage{graphicx}
\usepackage{dcolumn}
\usepackage{bm}
\usepackage{multirow}

\pdfoutput=1

\begin{document}
\preprint{APS/123-QED}
\title{Evidence for M-theory based on fractal nearly tri-bimaximal neutrino mixing}
\author{Hui-Bin Qiu}
\email{Corresponding author: huibinqiu@ncu.edu.cn}
\affiliation{Department of Physics, Nanchang University, JiangXi, Nanchang 330031, China}

\date{\today}

\begin{abstract}

\noindent Developing a theory that can describe everything in the universe is of great interest, and is closely relevant to M-theory, neutrino oscillation and charge-parity (CP) violation. Although M-theory is claimed as a grand unified theory, it has not been tested by any direct experiment. Here we show that existing neutrino oscillation experimental data supports one kind of high dimensional unified theory, such as M-theory. We propose a generalization of the tri-bimaximal neutrino mixing ansatz, and we find that the latest neutrino oscillation experimental data constraints dimension in a range between 10.46 and 12.93 containing 11, which is an important prediction of M-theory. This ansatz naturally incorporates the fractal feature of the universe and leptonic CP violation into the resultant scenario of \textit{fractal} nearly tri-bimaximal flavor mixing. We also analyze the consequences of this new ansatze on the latest experimental data of neutrino oscillations, and this theory matches the experimental data. Furthermore, an approach to construct lepton mass matrices in fractal universe under permutation symmetry is discussed. The proposed theory opens an unexpected window on the physics beyond the Standard Model.
\begin{description}
\item[PACS numbers]{12.60.-i, 11.25.Yb, 14.60.Pq, 05.45.Df}
\end{description}
\end{abstract}
\maketitle


\emph{Introduction.}-- M-theory, one of the most promising theories beyond the Standard Model, is suffering from pseudoscience questions \cite{Ellis:2014} because of the lack of direct experimental evidence, and causes wide discussions \cite{Castelvecchi:2015}. Recently acquired Neutrino oscillation experimental data might provide promising chances to either support or decline the M-theory. However, the relationship between neutrino oscillation and M-theory has not been fully established yet. This is because the dimensions of these two theories are not identical. Neutrino theory is a low-dimensional theory while the M-theory is 11-dimensional \cite{Hull:1995,Witten:1995}. Usually, the high-energy M-theory has to be shrunk to 4 dimensions, forming a low-energy theory to match the experimental data such as those from Large Hadron Collider (LHC). However, none of these predictions have been supported by the LHC data yet because the low-dimension M-theory has not been completely developed.

Here, we expand the neutrino oscillation theory to 11-dimension using nonextensive statistics \cite{Tsallis:2009,Borges:1998}. This method has succeeded in many fields such as generalizations of relativistic and quantum equations \cite{Nobre:2011}, transverse momenta distributions at
LHC experiments \cite{Abelev:2013}, dissipative optical lattices \cite{Lutz:2013}, plasmas \cite{Qiu:2016}, etc (see http://tsallis.cat.cbpf.br/biblio.htm, for a regularly updated bibliography). Nonextensive statistics is based on the fractal principle \cite{Tsallis:1988}. We bring it to modify the tri-bimaximal neutrino mixing pattern, which allows to incorporate CP violation and the fractal feature of the universe into the resultant scenario of fractal nearly tri-bimaximal flavor mixing. Results show that the dimension of a neutrino system using the nonextensive statistics is located between 10.46 and 12.93, which well covers the 11 predicted by the M-theory.

\begin{figure}[htbp]
\centerline{\includegraphics[width=86mm,height=3in]{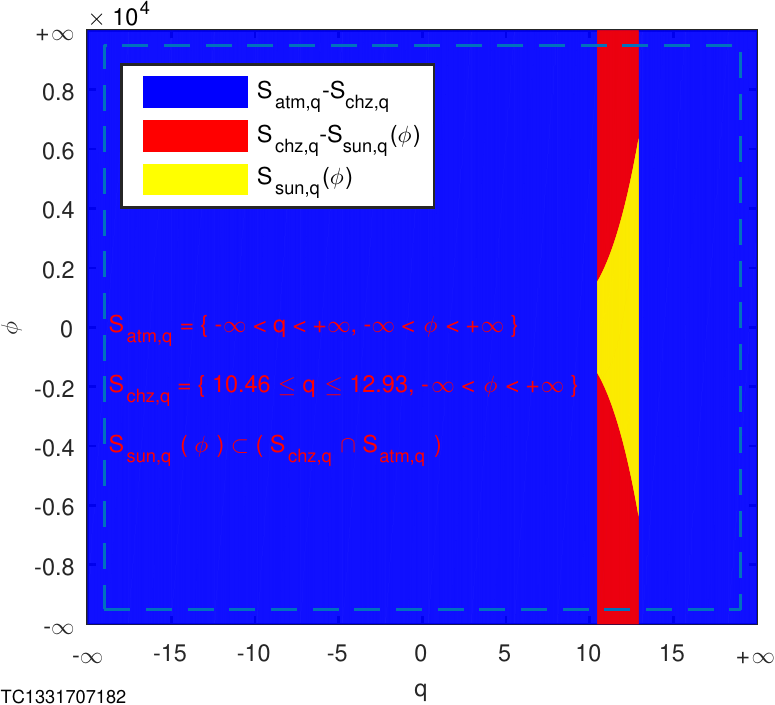}}
\caption{The space of dimension $q$ and phase $\phi $. The figure expresses
three sets and relationship among them: $
 S_{chz,q} =\left\{ {10.46\le q\le 12.93,-\infty <\phi <+\infty } \right\},
 S_{atm,q} =\left\{ {-\infty <q<+\infty ,-\infty <\phi <+\infty }
\right\}\supset S_{chz,q}, 
 S_{sun,q} \left( \phi \right)\subset S_{chz,q}$, 
where, the range of $q$ in set $S_{chz,q} $ is obtained based
on experimental data $\sin ^22\theta _{chz} =\left( {8.5\pm 0.5}
\right)\times 10^{-2}$, and $\phi $ is not limited now, so it can take any
real number. $S_{atm,q} $ is decided by experimental data $\sin ^2\left(
{2\theta _{23} } \right)=0.999_{-0.018}^{+0.001} $ for normal mass hierarchy
and $\sin ^2\left( {2\theta _{23} }
\right)=\mbox{1}.\mbox{000}_{-0.017}^{+0.000} $ for inverted
mass hierarchy, and after checked there is $S_{atm,q} \supset S_{chz,q} $. In
fact, $q$ in $S_{atm,q} $ can take all real number due to the fact that the
experimental upper and lower limits of $\sin ^2\left( {2\theta _{23} }
\right)$ are automatically satisfied. For upper limit, $\sin ^22\theta _{atm}
=1-s^4\le 1,\forall q\in \mathbb{R}$, and for lower limit one has $\sin ^22\theta
_{atm} \ge 0.99997675$, which satisfies the lower limit. At this moment $\phi
$ is also not limited, so it can take any real number too On the basis of
meet the above conditions $S_{sun,q} \left( \phi \right)$ is decided by
experimental data $\sin ^2\left( {2\theta _{12} } \right)=0.846\pm 0.021$,
So $S_{sun,q} \left( \phi \right)\subset \left( {S_{chz,q} \cap S_{atm,q} }
\right)$. Therefore, the value of $q$ in $S_{sun,q} \left( \phi \right)$ is
$\left\{ {10.46\le q\le 12.93} \right\}$, and the value of $\phi $ is
decided by experimental data $\sin ^2\left( {2\theta _{12} }
\right)=0.846\pm 0.021$, in fact only by $\sin ^22\theta _{sun} \le 0.867$.
$S_{sun,q} \left( \phi \right)$ is expressed by the yellow area in the figure.}
\label{fig1}
\end{figure}

%
\emph{Constraints on dimension and mixing factors.}--
In order to obtain the dimension range of neutrino system, we analyze the
latest neutrino oscillation experimental data with fractal nearly
tri-bimaximal neutrino mixing theory (see Appendixes).
In detail, adopting theoretical formula (see Eq. (15) in Appendixes) $\sin ^22\theta _{chz} =1-c^4$, in which $c\equiv \cos _q \theta $, combining with experimental
data \cite{Olive:2014} $\sin ^22\theta _{chz} =\left( {8.5\pm 0.5}
\right)\times 10^{-2}$, we obtain the allowed range of space-time dimension
(there is an intimate relation $q=d_f $ between $q$ and fractal dimension
$d_f $ when the Euclidean dimension is one \cite{Tsallis:1995}):
$\mbox{10.46}\le q\le 12.93$. This moment, theoretical formula and
experimental data have no limit on $\phi $ which is the source of leptonic
CP violation in neutrino oscillations, so there is a set $S_{chz,q} =\left\{
{10.46\le q\le 12.93,-\infty <\phi <+\infty } \right\}$, which can be
expressed in Fig. \ref{fig1} with the red strip area.
For sake of seeing the limit of theoretical formula (see Eq. (15) in Appendixes) $\sin ^22\theta _{atm} =1-s^4$, in which $s\equiv \sin _q \theta$, and experimental
data \cite{Olive:2014} $\sin ^2\left( {2\theta _{23} }
\right)=0.999_{-0.018}^{+0.001} $ for normal mass hierarchy and $\sin
^2\left( {2\theta _{23} } \right)=1.000_{-0.017}^{+0.000} $ for inverted
mass hierarchy on the range of $q$ and $\phi $, we do the corresponding
calculation and find that $q$ can take any real number, namely, $-\infty
<q<+\infty $, and this moment, theoretical formula and experimental data
also have no limit on $\phi $. So, there is a set $S_{atm,q} =\left\{
{-\infty <q<+\infty ,-\infty <\phi <+\infty } \right\}$ which can be
expressed in Fig. 1 with the blue strip area. There is relationship
$S_{atm,q} \supset S_{chz,q} $, seeing Fig. 1. Take the intersection of
these two sets we conclude that the range of space-time dimension that our
theory combined with the latest neutrino oscillation experimental data
allowed is between 10.46 and 12.93 containing 11, which is an important
prediction of M-theory. We can also see that the allowed range of space-time
dimension will be further restricted with the improvement of the
experimental accuracy. The neutrino oscillation experimental data becomes
the first evidence of M-theory, which will effectively eliminate the
people's question to M-theory \cite{Ellis:2014}.

With the purpose of obtaining the range of $\phi $, we adopt theoretical formula
(see Eq. (15) in Appendixes) $\sin ^22\theta _{sun} =\frac{8}{9}\left(
{1-\frac{3}{4}s^2-sc\cos _q \phi +\frac{3}{2}s^3c\cos _q \phi
-2s^2c^2\cos _q ^2\phi } \right)$ to analyze the experimental
data \cite{Olive:2014} $\sin ^2\left( {2\theta _{12} } \right)=0.846\pm
0.021$. We using numerical calculation find that the range of $\phi $ is
depending on parameter $q$. The top and bottom limit of $\phi $ under the
typical $q$ values are in Table~\ref{tab1}.
\begin{table}[b]
\caption{\label{tab1}%
The range of $\phi$ and strength of CP or T violation. $q=1$ is the ideal one-dimensional case; $q=10.46$ and $12.93$ are
dimension lower and upper limits allowed by existing neutrino oscillation
experimental data, respectively; $q=11$ is the prediction of M-theory. After
the dimension increased, the range of phase spanned from $0.49\le \phi
_{q=1} \le 1.27$ to $-2162.81\le \phi _{q=11} \le 2162.81$, increasing
3 orders; The predicted strength of CP violation is $-0.0011\le J_{q=11} \le
0.0011$, which can be determined by the T- or CP-violating asymmetry in a long-baseline neutrino oscillation
experiment.
}
\begin{ruledtabular}
\begin{tabular}{ddddd}
\multicolumn{1}{c}{\textrm{$q$}}&
\multicolumn{1}{c}{\textrm{$\phi_{\min}$}}&
\multicolumn{1}{c}{\textrm{$\phi_{\max}$}}&
\multicolumn{1}{c}{\textrm{$J_{\min}$}}&
\multicolumn{1}{c}{\textrm{$J_{\max}$}}\\
\colrule
1.00&
0.49&
1.27&
0.0054&
0.0110 \\
10.46&
-1537.79&
1537.79&
-0.0012&
0.0012 \\
11.00&
-2162.81&
2162.81&
-0.0011&
0.0011 \\
12.93&
-6372.47&
6372.47&
-0.0009&
0.0009 \\
\end{tabular}
\end{ruledtabular}
\end{table}
The $\phi $ set under the $q$ that
allowed by all theoretical formula and experimental data, $S_{sun,q} \left(
\phi \right)$, can be expresses with the yellow area in Fig. 1, and there is
relationship $S_{sun,q} \left( \phi \right)\subset S_{chz,q} $.

In conclusion, the set of $q$ and $\phi $ allowed by theoretical formula
and experimental data is the intersection of sets $S_{chz,q} $, $S_{atm,q} $
and $S_{sun,q} \left( \phi \right)$, namely, $S_{sun,q} \left( \phi
\right)$, i.e. the yellow area in Fig. 1.

%
%
\emph{Change on mixing factors.}--
Next, we investigate the change of range of $\phi $ after the dimension
increased with the theoretical formula (see Eq. (15) in Appendixes)
$\sin ^22\theta _{sun} =\frac{8}{9}\left( {1-\frac{3}{4}s^2-sc\cos _q
\phi +\frac{3}{2}s^3c\cos _q \phi -2s^2c^2\cos _q ^2\phi } \right)$
under the cases of $q=1$ and $10.46\le q\le 12.93$, respectively. From Table
1 and Fig. 2 we find that when $q=1$, $\mbox{0.49}\le \phi _{q=1} \le
1.27$ the order of magnitude is 1; but when $10.46\le q\le 12.93$, the
range of $\phi $ increases with the increase of $q$ namely, from
$-1537.79\le \phi _{q=10.46} \le 1537.79$ to $-\mbox{6372.47}\le
\phi _{q=12.93} \le \mbox{6372.47}$ with the order of
magnitude $\mbox{10}^3$. So, the order of magnitude of $\phi $
range increases 3 order after the dimension increased, which eliminates the
question of small range of $\phi $ values. Specifically, when $q=11$,
$-\mbox{2162.81}\le \phi _{q=11} \le \mbox{2162.81}$.
\begin{figure}[htbp]
\center{\includegraphics[width=86mm,height=3in]{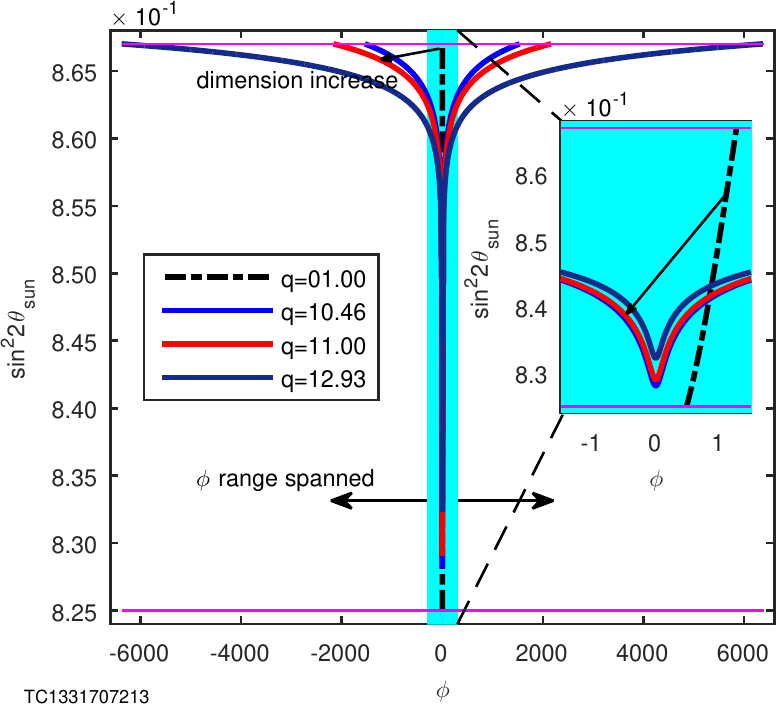}}
\caption{The mixing factors $\sin ^22\theta _{sun} $ against parameter
$\phi$ under different values of $q$ in fractal nearly tri-bimaximal
neutrino mixing patterns and the top and bottom limits of experimental data.
The two horizontal magenta lines are the top and bottom limits of
experimental data $\sin ^2\left( {2\theta _{12} } \right)=0.846\pm 0.021$.
Black dotted line is the limit case of $q\to 1$, and that time the experimental
data $\sin ^2\left( {2\theta _{12} } \right)=0.846\pm 0.021$ limits
${0.49}\le \phi _{q=1} \le 1.27$. Because the experimental
data $\sin ^22\theta _{chz} =\left( {8.5\pm 0.5} \right)\times
10^{-2}$ limits $10.46\le q\le 12.93$, the line of $q=1$ is not true. The
blue solid line is the experimental lower limit case ($q=10.46)$, and that
time the upper limit of experimental data $\sin ^2\left( {2\theta _{12} }
\right)=0.846\pm 0.021$ limits $-{1537.79}\le \phi _{q=10.46} \le
1537.79$. The purple solid line is the experimental upper limit case
($q=12.93)$, and that time the upper limit of experimental data $\sin
^2\left( {2\theta _{12} } \right)=0.846\pm 0.021$ limits $-{6372.47}\le
\phi _{q=12.93} \le {6372.47}$. The red solid line is the
M-theory predicted case ($q=11)$, that time the upper limit of experimental
data $\sin ^2\left( {2\theta _{12} } \right)=0.846\pm 0.021$ limits
$-{2162.81}\le \phi _{q=11} \le {2162.81}$. To
facilitate observing details, subgraph is the full figure's part of $-1.5\le
\phi \le 1.5$. The figure shows the order of magnitude of $\phi $ range
increases 3 orders after the dimension increased.}
\label{fig2}
\end{figure}

%
%
\emph{Prediction on CP violation.}--
To examine the theory proposed in this paper, we give a prediction
of the strength of CP or T violation in neutrino oscillations.
No matter whether neutrinos are Dirac or Majorana particles, the
strength of CP or T violation in neutrino oscillations is measured by a
universal parameter $J$ which is
defined as \cite{Jarlskog:1985}:
$Im\left( {V_{\alpha i} V_{\beta j} V_{\alpha j}^\ast V_{\beta i}^\ast }
\right)=J\sum\limits_{\gamma ,k} {\left( {\varepsilon _{\alpha \beta \gamma
} \varepsilon _{ijk} } \right)}$,
in which the Greek subscripts run over $\left( {e,\mu ,\tau } \right)$, and
the Latin subscripts run over $\left( {1, 2, 3} \right)$. Considering
the lepton mixing scenario proposed above, one has (see Eq. (18)
in Appendixes): $J=\frac{1}{6}sc\sin _q \phi \left( {c^2+s^2\rho _q^2
\left( \phi \right)} \right)$. The prediction of the strength of CP or T
violation in neutrino oscillations under typical $q$ values are in Table 1,
and especially, when $q=11$, $-\mbox{0.0011}\le \mbox{J}_{q=11} \le
\mbox{0.0011}$. Fig. 3 expresses the prediction intuitively.
The predicted strength of CP violation can be determined by the T-violating asymmetry between $\nu
_\mu \to \nu _e $ and $\nu _e \to \nu _\mu $ transitions or by the
CP-violating asymmetry between $\nu _\mu \to \nu _e $ and $\bar {\nu }_\mu
\to \bar {\nu }_e $ transitions in a long-baseline neutrino oscillation
experiment, when the terrestrial matter effects are under control or
insignificant.
\begin{figure}[htbp]
\centerline{\includegraphics[width=86mm,height=3in]{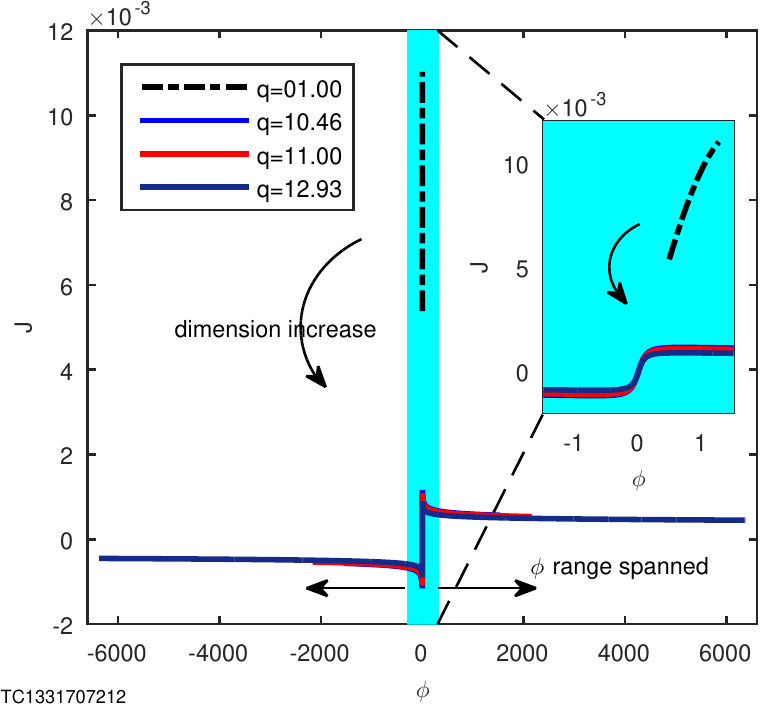}}
\caption{The strength of CP or T violation $J$ against parameter $\phi $
under different values of $q$ in fractal nearly tri-bimaximal neutrino
mixing patterns. Black dotted line is the limit case of $q\to 1$, and that time the experimental
data $\sin ^2\left( {2\theta _{12} } \right)=0.846\pm 0.021$ limits
${0.49}\le \phi _{q=1} \le 1.27$. Based on the figure as well as
the numerical calculations, one obtains $0.0054\le J_{q=1} \le 0.0110$. When
$q\to 1$, in theory, $\sin ^22\theta _{chz} =1-\cos ^4\theta =0.0096$, which
is not consistent with experimental data $\sin ^22\theta _{chz} =\left(
{{8.5}\pm {0.5}} \right)\times {10}^{-2}$, so the line of
$q=1$ is not true. The blue solid line is the experimental lower limit
case ($q=10.46)$, and that time the upper limit of experimental data $\sin
^2\left( {2\theta _{12} } \right)=0.846\pm 0.021$ limits $-{1537.79}\le
\phi _{q=10.46} \le 1537.79$. Based on the figure as well as the numerical
calculations, one obtains $-0.0012\le J_{q=10.46} \le 0.0012$. The purple solid
line is the experimental upper limit case ($q=12.93)$, and that time the
upper limit of experimental data $\sin ^2\left( {2\theta _{12} }
\right)=0.846\pm 0.021$ limits $-{6372.47}\le \phi
_{q=12.93} \le {6372.47}$. Based on the figure as well as
the numerical calculations, one obtains $-0.0009\le
J_{q=12.93} \le 0.0009$. The red solid line is the
M-theory predicted case ($q=11$), that time the upper limit of experimental
data $\sin ^2\left( {2\theta _{12} } \right)=0.846\pm 0.021$ limits
$-{2162.81}\le \phi _{q=11} \le {2162.81}$. Based on
the figure as well as the numerical calculations, one obtains prediction
$-0.0011\le J_{q=11} \le 0.0011$. To facilitate
observing details, subgraph is the full figure's part of $-1.5\le \phi \le
1.5$. The figure shows the predicted strength of CP violation is $-0.0011\le
J_{q=11} \le 0.0011$, which can be determined by the T- or CP-violating asymmetry in a long-baseline neutrino
oscillation experiment.}
\label{fig3}
\end{figure}

%
%
\emph{Further discussions and remarks.}--
Our findings reveal a strong association between neutrino oscillation and M-theory at the point of 11 dimensions of space-time. This would mean that the neutrino oscillation experiment is the initial robust evidence of M-theory, broking the spell that the M-theory has no experimental evidence, eliminating pseudoscience questions \cite{Ellis:2014}, and opening an unexpected window on the physics beyond the Standard Model. However, we should realize that in spite of the M-theory have part truth, but not completely developed yet, and there may be other way. Fractal theory and practice \cite{Mandelbrot:1982} have illuminated that the world is of fractal. The definition of fractal dimension is more universal than the one of Euclidean dimension. Euclid dimension is just a special case of fractal dimension, and there is intimate relation $q=d_f$ between $q$ and fractal dimension $d_f$ when the Euclidean dimension is one \cite{Tsallis:1995}. As the $q \to 1$ limit case of the fractal nearly tri-bimaximal neutrino mixing pattern under discussion, the nearly tri-bimaximal neutrino mixing pattern, as Xing \cite{Zhi:2002} expected, serves as the leading-order approximation of a more complicated flavor mixing matrix  (see Eq. (12) in Appendixes), though its prediction on $\sin ^22\theta _{chz} $ is not consistent with the experimental data. Although existing neutrino oscillation experiment data limits the range of space-time dimension between 10.46 and 12.93 (see Fig. 1), the range of space-time dimension will be narrowed down with the increasing of experimental accuracy, and we expect an exclusion of 12 dimension. In addition, we find the order of magnitude of $\phi $ range increases 3 orders after the dimension increased (see Fig. 2). Moreover, this theory yields a prediction (see Fig. 3) which can be determined by the T-violating asymmetry
between $\nu _\mu \to \nu _e $ and $\nu _e \to \nu _\mu $ transitions or by the CP-violating asymmetry between $\nu _\mu \to \nu _e $ and $\bar {\nu
}_\mu \to \bar {\nu }_e $ transitions in a long-baseline neutrino oscillation experiment, when the terrestrial matter effects are under
control or insignificant. Note that our scenario predicts that $-\mbox{0.0011}\le J_{q=11} \le \mbox{0.0011}$, and when $\phi =0$,
$J_{q=11} =0$, namely, there is no CP violation. Therefore, our theory can be applied whether CP is violated or not.

Finally, let us remark that the fractal nearly tri-bimaximal mixing pattern and its possible extensions require some peculiar flavor symmetries to be imposed on the charged lepton and neutrino mass matrices. It is likely that the fractal nearly tri-bimaximal neutrino mixing pattern under discussion serves as the more complicated flavor mixing matrix that scientists are looking for \cite{Zhi:2002}, and one of the nearly tri-bimaximal neutrino mixing patterns is its leading-order approximation. We expect that more delicate neutrino oscillation experiments in the near future will be able to verify the fractal nearly tri-bimaximal mixing pattern, from which one may get some insight into the underlying flavor symmetry and its breaking mechanism responsible for the origin of both lepton masses and leptonic CP violation.


\vspace{1em}

\noindent

\section*{Appendixes}
\noindent
The mixing factors of solar, atmospheric and
CHOOZ neutrino oscillations read:
\begin{equation}
\label{eq1}
\begin{array}{l}
 \sin ^22\theta _{sun} =4\left| {V_{e1} } \right|^2\left| {V_{e\mbox{2}} }
\right|^2, \\
 \sin ^22\theta _{atm} =4\left| {V_{\mu 3} } \right|^2\left( {1-\left|
{V_{\mu 3} } \right|^2} \right),\\
 \sin ^22\theta _{chz} =4\left| {V_{e3} } \right|^2\left( {1-\left| {V_{e3}
} \right|^2} \right). \\
 \end{array}
\end{equation}

\vspace{1em}

\noindent
\subsection*{\textbf{A}. Fractal nearly tri-bimaximal neutrino mixing}

The tri-bimaximal neutrino mixing pattern $U_v =V_0 $ can be
constructed from the product of two modified Euler rotation matrices:

\begin{equation}
\label{eq5}
\begin{array}{l}
 R_{12} \left( {\theta _x } \right)=\left( {{\begin{array}{*{20}c}
 {c_x } \hfill & {s_x } \hfill & 0 \hfill \\
 {-s_x } \hfill & {c_x } \hfill & 0 \hfill \\
 0 \hfill & 0 \hfill & 1 \hfill \\
\end{array} }} \right), \\
 R_{23} \left( {\theta _y } \right)=\left( {{\begin{array}{*{20}c}
 1 \hfill & 0 \hfill & 0 \hfill \\
 0 \hfill & {c_y } \hfill & {s_y } \hfill \\
 0 \hfill & {-s_y } \hfill & {c_y } \hfill \\
\end{array} }} \right), \\
 \end{array}
\end{equation}
where $s_x \equiv \sin _q \theta _x ,c_y \equiv \cos _q \theta _y $, and so
on. Functions $\sin _q u$ and $\cos _q u$ can be defined with $\exp _q
\left( u \right)$ which is the one-dimensional q-exponential function that
naturally emerges in nonextensive statistics \cite{Tsallis:2009} spawned by fractal thought \cite{Tsallis:1988}. For a pure
imaginary $iu$, one defines $\exp _q \left( {iu} \right)$ as the principal
value of

\begin{equation}
\label{eq6}
\begin{array}{l}
 \exp _q \left( {iu} \right)=\left[ {1+\left( {1-q} \right)iu} \right]^{1
\mathord{\left/ {\vphantom {1 {\left( {1-q} \right)}}} \right.
\kern-\nulldelimiterspace} {\left( {1-q} \right)}}, \\
 \exp _1 \left( {iu} \right)\equiv \exp \left( {iu} \right). 
 \end{array}
\end{equation}
The above function satisfies \cite{Borges:1998}:
\begin{equation}
\label{eq7}
\exp _q \left( {\pm iu} \right)=\cos _q \left( u \right)\pm i\sin _q \left(
u \right),
\end{equation}
\begin{equation}
\label{eq8}
\cos _q \left( u \right)=\rho _q \left( u \right)\cos \left\{
{\frac{1}{q-1}\arctan \left[ {\left( {q-1} \right)u} \right]} \right\},
\end{equation}
\begin{equation}
\label{eq9}
\sin _q \left( u \right)=\rho _q \left( u \right)\sin \left\{
{\frac{1}{q-1}\arctan \left[ {\left( {q-1} \right)u} \right]} \right\},
\end{equation}
\begin{equation}
\label{eq10}
\rho _q \left( u \right)=\left[ {1+\left( {1-q} \right)^2u^2} \right]^{1
\mathord{\left/ {\vphantom {1 {\left[ {2\left( {1-q} \right)} \right]}}}
\right. \kern-\nulldelimiterspace} {\left[ {2\left( {1-q} \right)}
\right]}},
\end{equation}
\begin{equation}
\label{eq11}
\exp _q \left( {iu} \right)\exp _q \left( {-iu} \right)=\cos _q^2 \left( u
\right)+\sin _q^2 \left( u \right)=\rho _q^2 \left( u \right).
\end{equation}
Note that $\exp _q \left[ {i\left( {u_1 +u_2 } \right)} \right]\ne \exp _q
\left( {iu_1 } \right)\exp _q \left( {iu_2 } \right)$ for $q\ne
1$\cite{Tsallis:2009}. Then we obtain:
\begin{equation}
\label{eq12}
\begin{array}{l}
 V_0 =R_{23} \left( {\theta _y } \right)\otimes R_{12} \left( {\theta
_x } \right) \\
 =\left( {{\begin{array}{*{20}c}
 {c_x } \hfill & {s_x } \hfill & 0 \hfill \\
 {-s_x c_y } \hfill & {c_x c_y } \hfill & {s_y } \hfill \\
 {s_x s_y } \hfill & {-s_y c_x } \hfill & {c_y } \hfill \\
\end{array} }} \right) .\\
 \end{array}
\end{equation}
The general form of
the corresponding neutrino mass matrix $M_\nu $ is
\begin{widetext}
\begin{equation}
\label{eq13}
\begin{array}{l}
 M_\nu =V_0 \left( {{\begin{array}{*{20}c}
 {m_1 } \hfill & 0 \hfill & 0 \hfill \\
 0 \hfill & {m_2 } \hfill & 0 \hfill \\
 0 \hfill & 0 \hfill & {m_3 } \hfill \\
\end{array} }} \right)V_0^T \\
 =\left( {{\begin{array}{*{20}c}
 {c_x^2 m_1 +s_x^2 m_2 } \hfill & {-c_x c_y s_x \left( {m_1 -m_2 } \right)}
\hfill & {c_x s_x s_y \left( {m_1 -m_2 } \right)} \hfill \\
 {-\!c_x c_y s_x\! \left(\! {m_1\! -\!m_2 }\! \right)} \hfill & {c_y^2 s_x^2 m_1 \!+\!c_x^2
c_y^2 m_2\! +\!s_y^2 m_3 } \hfill & {\!-c_y s_y\! \left(\! {s_x^2 m_1 \!+\!c_x^2 m_2 \!-\!m_3
}\! \right)\!} \hfill \\
 {c_x s_x s_y \left( {m_1\! -\!m_2 } \right)} \hfill & {-\!c_y s_y\! \left( \!{s_x^2
m_1 \!+\!c_x^2 m_2 \!-\!m_3 }\! \right)\!} \hfill & {s_x^2 s_y^2 m_1 \!+\!c_x^2 s_y^2 m_2\!
+\!c_y^2 m_3 } \hfill \\
\end{array} }} \right). \\
 \end{array}
\end{equation}
\end{widetext}
Taking $q=1$, $\theta _x =\arctan \left( {1 \mathord{\left/ {\vphantom {1
{\sqrt 2 }}} \right. \kern-\nulldelimiterspace} {\sqrt 2 }} \right)\approx
35.3^\circ$ and $\theta _y =45^\circ$, the results in usual
space-time are reproduced\cite{Zhi:2002}.

To make CP violation and the fractal feature of the universe
be naturally incorporated into $V$, we adopt the following complex rotation matrices:
\begin{equation}
\label{eq17}
R_{12} \left( {\theta ,\phi } \right)=\left( {{\begin{array}{*{20}c}
 c \hfill & {se_q^{i\phi } } \hfill & 0 \hfill \\
 {-se_q^{-i\phi } } \hfill & c \hfill & 0 \hfill \\
 0 \hfill & 0 \hfill & 1 \hfill \\
\end{array} }} \right),
\end{equation}
where $c\equiv \cos _q \theta $, $s\equiv \sin _q \theta$, and $e_q^{i\phi }
=\exp _q \left( {i\phi } \right)$. In this case, we obtain the lepton
flavor mixing of the following pattern:
\begin{equation}
\label{eq18}
\begin{array}{l}
 V=R_{12}^\dag \left( {\theta ,\phi } \right)\otimes V_0 \\
 =\left( {{\begin{array}{*{20}c}
 {\frac{1}{\sqrt 6 }\left( {2c+se_q^{i\phi } } \right)} \hfill &
{\frac{1}{\sqrt 3 }\left( {c-se_q^{i\phi } } \right)} \hfill &
{-\frac{1}{\sqrt 2 }se_q^{i\phi } } \hfill \\
 {-\frac{1}{\sqrt 6 }\left( {c-2se_q^{-i\phi } } \right)} \hfill &
{\frac{1}{\sqrt 3 }\left( {c+se_q^{-i\phi } } \right)} \hfill &
{\frac{1}{\sqrt 2 }c} \hfill \\
 {\frac{1}{\sqrt 6 }} \hfill & {-\frac{1}{\sqrt 3 }} \hfill &
{\frac{1}{\sqrt 2 }} \hfill \\
\end{array} }} \right). \\
 \end{array}
\end{equation}
$V$ represents a fractal nearly tri-bimaximal flavor
mixing scenario, if the rotation angle $\theta $ is small. The
parameter $\phi $ in $V$ are the source of leptonic CP violation
in neutrino oscillations.

\vspace{1em}

\noindent
\subsection*{\textbf{B}. Constraints on dimension, mixing factors and CP violation}

A proper texture of $M_l $ which may lead to the flavor mixing
pattern $V$ is
\begin{equation}
\label{eq19}
M_l =\left( {{\begin{array}{*{20}c}
 0 \hfill & {C_l } \hfill & 0 \hfill \\
 {C_l^\ast } \hfill & {B_l } \hfill & 0 \hfill \\
 0 \hfill & 0 \hfill & {A_l } \hfill \\
\end{array} }} \right),
\end{equation}
where $A_l =m_\tau$, $B_l =m_\mu -m_e $, and $C_l =\sqrt {m_e m_\mu }
e_q^{i\phi } $. Then the mixing angle $\theta $ in $V$ reads
\begin{equation}
\label{eq20}
\tan _q \left( \theta \right)=\frac{\sin _q \theta }{\cos _q \theta }=\sqrt
{\frac{m_e }{m_\mu }} .
\end{equation}
It is easy to prove that when $q\to 1$, the results in usual space-time are
recovered, namely \cite{Zhi:2002},
$ C_l =\sqrt {m_e m_\mu } e^{i\phi }$, $ \tan \left( {2\theta } \right)=2\frac{\sqrt {m_e m_\mu } }{m_\mu -m_e }$.

In the next step we calculate the mixing factors of solar, atmospheric and reactor
neutrino oscillations.
According to this theory, one obtains
\begin{widetext}
\begin{equation}
\label{eq22}
\begin{array}{l}
 \sin ^22\theta _{sun} =\frac{8}{9}\left( {1-\frac{3}{4}s^2-sc\cos _q
\phi +\frac{3}{2}s^3c\cos _q \phi -2s^2c^2\cos _q ^2\phi } \right),
\\
 \sin ^22\theta _{atm} =1-s^4, \\
 \sin ^22\theta _{chz} =1-c^4. \\
 \end{array}
\end{equation}
Note when $q\to 1$, the results in usual space-time are
recovered \cite{Zhi:2002}:
\begin{equation}
\label{eq23}
\begin{array}{l}
 \sin ^22\theta _{sun}\! =\!\frac{8}{9}\!\left(\!{1\!-\!\frac{3}{4}\!\sin ^2\!\theta\! -\!\sin\!
\theta \cos\! \theta \cos \!\phi \! +\!\frac{3}{2}\!\sin ^3\!\theta \cos \!\theta \cos\!
\phi\! -\!2\!\sin ^2\!\theta \cos ^2\!\theta \cos ^2\!\phi } \right)\!, \\
 \sin ^22\theta _{atm} =1-\sin ^4\theta , \\
 \sin ^22\theta _{chz} =1-\cos ^4\theta . \\
 \end{array}
\end{equation}
\end{widetext}

In this scenario, adopting experimental data \cite{Olive:2014} $\sin
^22\theta _{chz}=\left( {\mbox{8.5}\pm \mbox{0.5}} \right)\times
10^{-2}$, one obtains $\mbox{10.46}\le q\le
\mbox{12.93}$; thus there is $0.999987\le \sin ^22\theta
_{atm} \le 0.99999$, which is highly consistent with the
experimental data \cite{Olive:2014}: $\sin ^2\left( {2\theta _{23} }\right) \\
=0.999_{-0.018}^{+0.001} $ for normal mass hierarchy and $\sin
^2\left( {2\theta _{23} } \right)=1.000_{-0.017}^{+0.000} $ for inverted
mass hierarchy; in addition, to make $\sin ^22\theta _{sun} \le 0.867$ to
accord with the experimental data $\sin ^2\left( {2\theta _{12} }
\right)=0.846\pm 0.021$, one needs only $-1537.79\le \phi
_{q=10.46} \le 1537.79$ or $-6372.47\le \phi
_{q=12.93} \le 6372.47$, which are much better than the usual
space-time case ($0.49 \le \phi _{q=1} \le 1.27)$.

Additionally, given that $q$ is close to 11 and the intimate relation $q=d_f $
between $q$ and fractal dimension $d_f $ when the Euclidean dimension is
one \cite{Tsallis:1995}, we assume $q=11$, then this scenario gives the
predicted values of $\sin ^22\theta _{chz}=0.082456$ and $\sin
^22\theta _{atm} =0.999987$ which amazingly fit in with
the current data \cite{Olive:2014} $\sin ^2\left( {2\theta _{13} }
\right)=\left( {8.5\pm 0.5} \right)\times 10^{-2}$ and $\sin ^2\left(
{2\theta _{23} } \right)=0.999_{-0.018}^{+0.001} $ for normal mass hierarchy
($\sin ^2\left( {2\theta _{23} } \right)=1.000_{-0.017}^{+0.000} $ for
inverted mass hierarchy), respectively; the range of parameter
$-2162.81\le \phi _{q=11} \le 2162.81$ limited by current
data $\sin ^2\left( {2\theta _{12} } \right)=0.846\pm 0.021$ is also much
better than that in usual space-time ($0.49 \le \phi _{q=1}
\le 1.27)$. According to the calculations above, we come to the following conclusions: i)
the universe is fractal with high dimension; ii) some high dimensional
space-time theories, such as M-theory, can be in line with
expectations.
A numerical illustration of $\sin ^22\theta _{sun}$ as the function of $q$
and $\phi $ is shown in Fig. 2, where the two horizontal lines are
the top and bottom limits of experimental data. As can be seen from the figure, in $\phi = 0$ case, $\sin ^22\theta _{sun}$ very sensitively dependent on $\phi $.

The strength of CP or T violation in neutrino oscillations is measured by a
universal parameter $J$ which is
defined as \cite{Jarlskog:1985}:
\begin{equation}
\label{eq24}
Im\left( {V_{\alpha i} V_{\beta j} V_{\alpha j}^\ast V_{\beta i}^\ast }
\right)=J\sum\limits_{\gamma ,k} {\left( {\varepsilon _{\alpha \beta \gamma
} \varepsilon _{ijk} } \right)}.
\end{equation}
Considering
the lepton mixing scenario proposed above, one has
\begin{equation}
\label{eq25}
J=\frac{1}{6}sc\sin _q \phi \left( {c^2+s^2\rho _q^2 \left( \phi
\right)} \right).
\end{equation}
Obviously, when $q\to 1$, the result in usual space-time is
recovered \cite{Zhi:2002}:
\begin{equation}
\label{eq26}
J=\frac{1}{6}sc\sin \phi.
\end{equation}
Based on Figs. 2 and 3 as well as the numerical calculations, one obtains the
table~\ref{tab1}.

The strength of CP or T violation $J$ in fractal nearly tri-bimaximal
neutrino mixing patterns is predicted as:
$-{0.0011}\le J_{q=11} \le {0.0011}$.
The experimental data of strength of CP or T violation may limit the range
of parameter $\phi $, but unfortunately at present, there is no experimental
information on the Dirac and Majorana CP violation phases in the neutrino
mixing matrix is available \cite{Olive:2014}.


\begin{thebibliography}{16}
\expandafter\ifx\csname natexlab\endcsname\relax\def\natexlab#1{#1}\fi
\expandafter\ifx\csname bibnamefont\endcsname\relax
  \def\bibnamefont#1{#1}\fi
\expandafter\ifx\csname bibfnamefont\endcsname\relax
  \def\bibfnamefont#1{#1}\fi
\expandafter\ifx\csname citenamefont\endcsname\relax
  \def\citenamefont#1{#1}\fi
\expandafter\ifx\csname url\endcsname\relax
  \def\url#1{\texttt{#1}}\fi
\expandafter\ifx\csname urlprefix\endcsname\relax\def\urlprefix{URL }\fi
\providecommand{\bibinfo}[2]{#2}
\providecommand{\eprint}[2][]{\url{#2}}

\bibitem[{\citenamefont{Ellis and Silk}(2014)}]{Ellis:2014}
\bibinfo{author}{\bibfnamefont{G.}~\bibnamefont{Ellis}} \bibnamefont{and}
  \bibinfo{author}{\bibfnamefont{J.}~\bibnamefont{Silk}},
  \bibinfo{journal}{Nature} \textbf{\bibinfo{volume}{516}},
  \bibinfo{pages}{321} (\bibinfo{year}{2014}).

\bibitem[{\citenamefont{Castelvecchi}(2015)}]{Castelvecchi:2015}
\bibinfo{author}{\bibfnamefont{D.}~\bibnamefont{Castelvecchi}},
  \bibinfo{journal}{Nature} \textbf{\bibinfo{volume}{528}},
  \bibinfo{pages}{446} (\bibinfo{year}{2015}).

\bibitem[{\citenamefont{Hull and Townsend}(1995)}]{Hull:1995}
\bibinfo{author}{\bibfnamefont{C.}~\bibnamefont{Hull}} \bibnamefont{and}
  \bibinfo{author}{\bibfnamefont{P.}~\bibnamefont{Townsend}},
  \bibinfo{journal}{Nuclear Physics B} \textbf{\bibinfo{volume}{438}},
  \bibinfo{pages}{109 } (\bibinfo{year}{1995}), ISSN \bibinfo{issn}{0550-3213}.

\bibitem[{\citenamefont{Witten}(1995)}]{Witten:1995}
\bibinfo{author}{\bibfnamefont{E.}~\bibnamefont{Witten}},
  \bibinfo{journal}{Nuclear Physics B} \textbf{\bibinfo{volume}{443}},
  \bibinfo{pages}{85 } (\bibinfo{year}{1995}), ISSN \bibinfo{issn}{0550-3213}.

\bibitem[{\citenamefont{Tsallis}(2009)}]{Tsallis:2009}
\bibinfo{author}{\bibfnamefont{C.}~\bibnamefont{Tsallis}},
  \emph{\bibinfo{title}{Introduction to nonextensive statistical mechanics :
  approaching a complex world}} (\bibinfo{publisher}{Springer},
  \bibinfo{year}{2009}).

\bibitem[{\citenamefont{Borges}(1998)}]{Borges:1998}
\bibinfo{author}{\bibfnamefont{E.~P.} \bibnamefont{Borges}},
  \bibinfo{journal}{Journal of Physics A: Mathematical and General}
  \textbf{\bibinfo{volume}{31}}, \bibinfo{pages}{5281} (\bibinfo{year}{1998}).

\bibitem[{\citenamefont{Nobre et~al.}(2011)\citenamefont{Nobre, Rego-Monteiro,
  and Tsallis}}]{Nobre:2011}
\bibinfo{author}{\bibfnamefont{F.~D.} \bibnamefont{Nobre}},
  \bibinfo{author}{\bibfnamefont{M.~A.} \bibnamefont{Rego-Monteiro}},
  \bibnamefont{and} \bibinfo{author}{\bibfnamefont{C.}~\bibnamefont{Tsallis}},
  \bibinfo{journal}{Phys. Rev. Lett.} \textbf{\bibinfo{volume}{106}},
  \bibinfo{pages}{140601} (\bibinfo{year}{2011}).

\bibitem[{\citenamefont{Abelev et~al.}(2013)\citenamefont{Abelev, Adam,
  Adamov\'a, Adare, Aggarwal, Aglieri~Rinella, Agnello, Agocs, Agostinelli,
  Ahammed et~al.}}]{Abelev:2013}
\bibinfo{author}{\bibfnamefont{B.}~\bibnamefont{Abelev}},
  \bibinfo{author}{\bibfnamefont{J.}~\bibnamefont{Adam}},
  \bibinfo{author}{\bibfnamefont{D.}~\bibnamefont{Adamov\'a}},
  \bibinfo{author}{\bibfnamefont{A.~M.} \bibnamefont{Adare}},
  \bibinfo{author}{\bibfnamefont{M.~M.} \bibnamefont{Aggarwal}},
  \bibinfo{author}{\bibfnamefont{G.}~\bibnamefont{Aglieri~Rinella}},
  \bibinfo{author}{\bibfnamefont{M.}~\bibnamefont{Agnello}},
  \bibinfo{author}{\bibfnamefont{A.~G.} \bibnamefont{Agocs}},
  \bibinfo{author}{\bibfnamefont{A.}~\bibnamefont{Agostinelli}},
  \bibinfo{author}{\bibfnamefont{Z.}~\bibnamefont{Ahammed}},
  \bibnamefont{et~al.} (\bibinfo{collaboration}{ALICE Collaboration}),
  \bibinfo{journal}{Phys. Rev. Lett.} \textbf{\bibinfo{volume}{111}},
  \bibinfo{pages}{222301} (\bibinfo{year}{2013}).

\bibitem[{\citenamefont{{Lutz} and {Renzoni}}(2013)}]{Lutz:2013}
\bibinfo{author}{\bibfnamefont{E.}~\bibnamefont{{Lutz}}} \bibnamefont{and}
  \bibinfo{author}{\bibfnamefont{F.}~\bibnamefont{{Renzoni}}},
  \bibinfo{journal}{Nature Physics} \textbf{\bibinfo{volume}{9}},
  \bibinfo{pages}{615} (\bibinfo{year}{2013}).

\bibitem[{\citenamefont{Qiu et~al.}(2016)\citenamefont{Qiu, Song, and
  Liu}}]{Qiu:2016}
\bibinfo{author}{\bibfnamefont{H.-B.} \bibnamefont{Qiu}},
  \bibinfo{author}{\bibfnamefont{H.-Y.} \bibnamefont{Song}}, \bibnamefont{and}
  \bibinfo{author}{\bibfnamefont{S.-B.} \bibnamefont{Liu}},
  \bibinfo{journal}{Physics of Plasmas} \textbf{\bibinfo{volume}{23}},
  \bibinfo{eid}{032101} (\bibinfo{year}{2016}).

\bibitem[{\citenamefont{Tsallis}(1988)}]{Tsallis:1988}
\bibinfo{author}{\bibfnamefont{C.}~\bibnamefont{Tsallis}},
  \bibinfo{journal}{Journal of Statistical Physics}
  \textbf{\bibinfo{volume}{52}}, \bibinfo{pages}{479} (\bibinfo{year}{1988}),
  ISSN \bibinfo{issn}{0022-4715}.

\bibitem[{\citenamefont{Olive and {Particle Data Group}}(2014)}]{Olive:2014}
\bibinfo{author}{\bibfnamefont{K.}~\bibnamefont{Olive}} \bibnamefont{and}
  \bibinfo{author}{\bibnamefont{{Particle Data Group}}},
  \bibinfo{journal}{Chinese Physics C} \textbf{\bibinfo{volume}{38}},
  \bibinfo{pages}{090001} (\bibinfo{year}{2014}).

\bibitem[{\citenamefont{Tsallis}(1995)}]{Tsallis:1995}
\bibinfo{author}{\bibfnamefont{C.}~\bibnamefont{Tsallis}},
  \bibinfo{journal}{Fractals} \textbf{\bibinfo{volume}{03}},
  \bibinfo{pages}{541} (\bibinfo{year}{1995}).

\bibitem[{\citenamefont{Jarlskog}(1985)}]{Jarlskog:1985}
\bibinfo{author}{\bibfnamefont{C.}~\bibnamefont{Jarlskog}},
  \bibinfo{journal}{Phys. Rev. Lett.} \textbf{\bibinfo{volume}{55}},
  \bibinfo{pages}{1039} (\bibinfo{year}{1985}).

\bibitem[{\citenamefont{Mandelbrot}(1982)}]{Mandelbrot:1982}
\bibinfo{author}{\bibfnamefont{B.}~\bibnamefont{Mandelbrot}},
  \emph{\bibinfo{title}{The fractal geometry of nature}}
  (\bibinfo{publisher}{W. H. Freeman and Co., San Francisco},
  \bibinfo{year}{1982}).

\bibitem[{\citenamefont{Xing}(2002)}]{Zhi:2002}
\bibinfo{author}{\bibfnamefont{Z.-Z.} \bibnamefont{Xing}},
  \bibinfo{journal}{Physics Letters B} \textbf{\bibinfo{volume}{533}},
  \bibinfo{pages}{85 } (\bibinfo{year}{2002}), ISSN \bibinfo{issn}{0370-2693}.

\end{thebibliography}

\end{document}